\documentclass[titlepage,prb,aps,groupedaddress,showpacs,byrevtex,preprint,12pt,nobibnotes]{revtex4}

\usepackage[dvips]{graphicx}%
\usepackage{amsmath}%

\begin{document}


\title{A model for luminescence of localized state ensemble}

\author{Q.~Li}
\affiliation{Department of Physics and HKU-CAS Joint Laboratory on New
Materials, The University of Hong Kong, Pokfulam Road, Hong Kong, China}

\author{S.~J.~Xu}
\thanks{Correspondence author. Electronic mail: sjxu@hkucc.hku.hk.}
\affiliation{Department of Physics and HKU-CAS Joint Laboratory on New
Materials, The University of Hong Kong, Pokfulam Road, Hong Kong,
China}

\author{M.~H.~Xie}
\affiliation{Department of Physics and HKU-CAS Joint Laboratory on New
Materials, The University of Hong Kong, Pokfulam Road, Hong Kong, China}

\author{S.~Y.~Tong}
\thanks{Present address: Department of Physics and Materials Science,
City University of Hong Kong, Kowloon, Hong Kong.}
\affiliation{Department of Physics and HKU-CAS Joint Laboratory on New
Materials, The University of Hong Kong, Pokfulam Road, Hong Kong, China}

\date{}



\begin{abstract}%

A distribution function for localized carriers,
$f(E,T)=\frac{1}{e^{(E-E_a)/k_BT}+\tau_{tr}/\tau_r}$, is proposed by
solving a rate equation, in which, electrical carriers' generation,
thermal escape, recapture and radiative recombination are taken into
account. Based on this distribution function, a model is developed for
luminescence from localized state ensemble with a Gaussian-type
density of states. The model reproduces quantitatively all the
anomalous temperature behaviors of localized state luminescence. It
reduces to the well-known band-tail and luminescence quenching models
under certain approximations.

\end{abstract}%


\maketitle



Carrier localization is a common phenomenon in many material systems
such as semiconductor alloys, quantum wells (QWs) and self-assembled
quantum dots (QDs). It has profound effects on electrical and optical
properties of the materials. It has been known for long that a number
of anomalous temperature-dependent luminescence behaviors are related
to carrier localization, including 1) the ``S-shaped'' temperature
dependence of the luminescence peak position
\cite{ChoYH1998-APL73:1370, EliseevPG1997-APL71:569,
  CaoXA2003-APL82:3614, KondowM1989-APL54:1760, XuZY1996-PRB54:11528}
and 2) a reduction of luminescence linewidth with increasing
temperature.\cite{XuZY1996-PRB54:11528, SanguinettiS1999-PRB60:8276}
Till now, no theory based on microscopic viewpoint is present, which
is capable of providing satisfactory explanations for all the
anomalous behaviors of the luminescence. In this work, we propose a
distribution function for localized carriers and develop a model for
the luminescence of localized state ensemble (LSE). The model suggests
that thermal redistribution of localized carriers within the localized
states is the cause of the anomalies in the temperature dependence of
the luminescence peak. It is further shown that the current model
reduces to the well-known band-tail model
\cite{EliseevPG1997-APL71:569} at high temperature and to the
luminescence quenching model of a two-level system \cite{CurieD1963}
when the distribution of localized states approaches a
$\delta$-function. The model is applied to quantitatively interpret
the luminescence data of two kinds of material systems, namely
InGaAs/GaAs self-assembled QDs and InGaN/GaN QDs. In the former system,
LSE naturally form due to size-distributed QDs while in the later, LSE
can form due to the In-rich clusters with broad size distribution.
The physical implication of $E_a-E_0$
and its effect on the temperature dependence of
luminescence peak position are revealed.



For a system with localized electronic states having the density of
states (DOS) $\rho(E)$, the rate of change of carriers' population
density $N(E,T)$ in the state at energy $E$ and temperature $T$ is
given by \cite{XuZY1998-SM23:381}
\begin{widetext}\begin{eqnarray}\label{Rate-eq}
 \frac{dN(E,T)}{dt}=G(E)+\frac{\gamma_cN'}{\Lambda}\rho(E)
 -\frac{N(E,T)}{\tau_{tr}}e^{(E-E_a)/k_BT}-\frac{N(E,T)}{\tau_r},
\end{eqnarray}\end{widetext}
where $E_a$\ is the energetic position of a delocalized state to which
the localized carriers thermally escape. The first term on the right,
$G(E)$, represents the rate of carrier generation, which is
proportional to $\rho(E)$ according to $G(E)=\kappa\cdot\rho(E)$,
\cite{XuZY1998-SM23:381} where $\kappa$ is a constant proportional to
absorption cross section and the number of incident excitation
photons. Note that the carrier generation includes both excitation of
carriers instantly in localized states and capture of carriers those
be excited outside. The second term on the right represents the number
of carriers re-captured by the localized states per unit time, in
which $\gamma_c$ is the recapture coefficient $N'$ is the total
number of carriers that are thermally activated away from the
localized states as given by
\begin{eqnarray}
 N'=\int_{-\infty}^{+\infty}\frac{N(E',T)}{\tau_{tr}}%
 e^{(E'-E_a)/k_BT}dE',
\end{eqnarray}
in which $1/\tau_{tr}$ is the escape rate of the localized
carrier. $\Lambda\!=\!\int_{-\infty}^{\infty}\rho(E')dE'$ is the total
number of localized states.
The third term on the right hand side of Eq.~(\ref{Rate-eq}) gives the
thermal escape rate of the localized carriers and the last term
describes the de-population rate of the carriers due to radiative
recombination. The term $1/\tau_r$ represents the rate of radiative
recombination. $\tau_{tr}$ and $\tau_r$ are assumed to be the same for
all the localized states. In Eq.~(\ref{Rate-eq}) the tunneling
transfer of carriers \cite{LubyshevDI1996-APL68:205} between QDs is
not taken into account because the non-resonant tunneling rate is
negligibly small.\cite{LuW2003-Nature} More detailed discussion
will be given later. The solution of Eq.~(\ref{Rate-eq}) under
steady-state condition ($dN/dt$=0) is
\begin{subequations}\label{solution}
\begin{eqnarray}\label{carrier-number}
  N(E,T)=A\cdot n(E,T),
\end{eqnarray}
where
\begin{eqnarray}
 A=\frac{\kappa\tau_{tr}}{(1-\gamma_c)
 +(\tau_{tr}/\tau_r)\gamma_c\xi_1/\Lambda},
 \label{amp}\\
 n(E,T)=\frac{\rho(E)}{e^{(E-E_a)/k_BT}+\tau_{tr}/\tau_r}.
 \label{distribution}
\end{eqnarray} 
\end{subequations}
In Eq.~(\ref{amp}),
$\xi_1=\int_{-\infty}^{+\infty}n(E',T)dE'$. Further expressing
$n(E,T)$ as $n(E,T)=f(E,T)\cdot\rho(E)$, then a distribution function
can be derived as
\begin{eqnarray}
 f(E,T)=\frac{1}{e^{(E-E_a)/k_BT}+\tau_{tr}/\tau_r}.
 \label{distribution2}%
\end{eqnarray}
Note that $n(E,T)$ essentially describes the shape of the luminescence
spectrum given by $N(E,T)/\tau_r$, because $A$ is a function of $T$
only. Such a quantity has previously been used to describe the
lineshape of phonon-assisted exciton luminescence peaks in some polar
semiconductors.\cite{PermogorovS-Excitons, CaswellN1981-PetroffY1975}



Considering a general case of a LSE system with Gaussian-type DOS,
which may result from, \textit{e.g.}, fluctuations of quantum dot size
or alloy composition:
\begin{eqnarray}
 \rho(E)=\rho_0e^{-(E-E_0)^2/2\sigma^2},
 \label{Gauss}
\end{eqnarray}
in above, $\rho_0$ and $E_0$ are the amplitude and peak energy
position, while $\sigma$ is the standard deviation of the
distribution. In Fig.~\ref{Fig-dist-funs}, the quantities of
$\rho(E)/\rho_0$ and $(\tau_{tr}/\tau_r)f(E,T)$ are plotted with
respect to $E$ for both $E_a>E_0$ and $E_a<E_0$. As will be shown in
the following, both cases exist in real physical systems as reported
in the literature.

As mentioned earlier, $n(E,T)$ represents the ``shape'' of the
luminescence spectrum and the luminescence peak position can be found
by setting $\partial n(E,T)/\partial E=0$. We found that when
\begin{eqnarray}
 E=E_0-x\cdot k_BT, \label{Energy}%
\end{eqnarray}
$n(E,T)$ reaches its maximum. The temperature-dependent coefficient
$x(T)$ can be obtained by numerically solving the following equation:
\begin{eqnarray}
 xe^x = \left[\left(\frac{\sigma}{k_BT}\right)^2-x\right]
 \left(\frac{\tau_{r}}{\tau_{tr}}\right)e^{(E_0-E_a)/k_BT},
 \label{sol-x}%
\end{eqnarray}
It is not difficult to see that Eq.~(\ref{sol-x}) has one and only one
solution for $0<x<(\sigma/k_BT)^2$.

It should be noted that the temperature dependence described by
Eq.~(\ref{Energy}) is only due to carriers' thermal redistribution
within the localized states. It is known that the bandgap of an
idealized semiconductor material without localized electronic states
is itself temperature dependent, which is usually described by the
Varshni empirical formula.\cite{VarshniYP1967-Physica34:149} After
taking into account this factor, the variation of the peak position of
luminescence from LSE is then given by
\begin{eqnarray}
 E(T)=E_0-\frac{\alpha T^2}{\Theta+T}-x\cdot k_BT,
 \label{Energy-Varshni}%
\end{eqnarray}
where $\alpha$ is the Varshni parameter and $\Theta$ the Debye
temperature of the material. For the cases where the thermal
redistribution of localized carriers is dominant in the temperature
range studied, the temperature dependence of the luminescence peak can
be well described with Eq.~(\ref{Energy}).\cite{LiQ2001-APL79:1810}

At high temperature region an approximated solution of
Eq.~(\ref{sol-x}) is found to be
$(\sigma/k_BT)^2$.\cite{LiQ2001-APL79:1810} Eq.~(\ref{Energy-Varshni})
then becomes
\begin{eqnarray}
 E(T)=E_0-\frac{\alpha T^2}{\Theta+T}-\frac{\sigma^2}{k_BT},
 \label{Energy-approx}%
\end{eqnarray}
which is simply the band-tail model proposed by Eliseev \textit{et
  al.} \cite{EliseevPG1997-APL71:569} So, the widely adopted band-tail
model can be viewed as an approximation of the current model under
high $T$ region.

Full width at half maximum height (FWHM) is another parameter for a
luminescence spectrum, which is also embedded in $n(E,T)$. The FWHM
$\Gamma_c(T)$ of $n(E,T)$ can be obtained by numerically solving
$n(E,T)=n(E_{pk},T)/2$. As will be shown below, the variation of
$\Gamma_c(T)$ with temperature exhibits a ``valley'', \textit{i.e.},
$\Gamma_c(T)$ decreases first and then increases with raising
temperature. Besides the variation in line width due to the thermal
redistribution of carriers within LSE, the broadening due to phonon
and impurity/imperfection scattering should be taken into
consideration. The effective FWHM of luminescence peak is determined
by making convolution of $n(E,T)$ and a Lorentzian function,
$[4E^2+(\Gamma_0+\Gamma_{ph})^2]^{-1}$. Here $\Gamma_0$ is due to
impurity/imperfection scattering and
$\Gamma_{ph}=\sigma_AT+\gamma_{LO}/[e^{\hbar\omega_{LO}/k_BT}-1]$ due
to phonon scattering.\cite{RudinS1990-PRB42:11218} In the RRS model
\cite{RudinS1990-PRB42:11218}, $\sigma_A$ and $\gamma_{LO}$ are the
acoustic- and optical-phonon coupling strength, respectively.

The integrated intensity of the luminescence spectrum is proportional
to the total number of localized carriers, \textit{i.e.},
\begin{eqnarray}
 I(T)\propto\int^{+\infty}_{-\infty}\!\!N(E',T)dE'=
 A\int^{+\infty}_{-\infty}\!\!n(E',T)dE'.
 \label{intensity}
\end{eqnarray}
Utilizing an integral approximation: \cite{LiQ2003-phdthesis}
\begin{eqnarray}
 \int_{-\infty}^{+\infty}\!\!\frac{e^{-x^2}}{1+e^{a(x+b)}}dx
 \approx\frac{\sqrt{\pi}}{1+e^{2.41b\sin{\theta}}},
 \label{int-approx}%
\end{eqnarray}
where $\theta=$arctan($a/2.41$), Eq.~(\ref{intensity}) can be derived
as \cite{LiQ2003-phdthesis}
\begin{widetext}\begin{eqnarray}
 I(T)\propto\left\{1+(1-\gamma_c)\cdot\exp\left[\frac{(E_0-E_a)+k_BT
 \cdot\ln(\tau_r/\tau_{tr})}
 {\sqrt{(k_BT)^2+2(\sigma/2.41)^2}}\right]\right\}^{-1}.
 \label{intensity-approx}%
\end{eqnarray}\end{widetext}
For $\sigma$=0 (\textit{i.e.}, a $\delta$-functional DOS), the above expression
is reduced to the well-known model describing thermal quenching of
luminescence for a two-level system.\cite{CurieD1963} Indeed, for the
case of $\sigma=0$, no thermal re-distribution takes place and the
system becomes essentially an equivalent two-level system. This fact
thus further validates the current model, which is more general. Note
that for the two-level system, $E_a-E_0$ is in effect the thermal
activation energy of the carriers.



Figure~\ref{Fig-N-ET} shows the calculated profiles for $N(E,T)$ at
different temperatures. The parameters used in the calculation are
$E_0$=1.185 eV, $E_a$=$E_0$+0.073 eV, $\sigma$=13 meV,
$\tau_{tr}/\tau_r$=0.027/250. From the figure, it is seen that the
typical anomalies in the temperature-dependent luminescence are
reproduced. Fig.~\ref{Fig-C258}(a) plots the peak positions of the
spectra as a function of temperature together with that predicted by
Varshni empirical formula using the Varshni parameter $\alpha$=0.48
meV/K and the Debye temperature $\Theta$=270 K. The sum of the two
contributions is given by the solid curve, which is seen to agree
excellently with the experimental data for an In$_{0.35}$Ga$_{0.65}$As
QD sample.\cite{LiQ2004} The QD density of the sample employed in the
present work is $\sim$5$\times$10$^{10}$ cm$^{-2}$. The average
distance between QDs is estimated as 20 nm. Other details of the
sample have been previously described
elsewhere. \cite{DuanRF2003-JJAP42:6314} The non-resonant tunneling
rate of carriers between QDs, which is estimated to be
$\sim$10$^5$~s$^{-1}$ using the Wentzel-Kramers-Brillouin
approximation, \cite{TackeuchiA2000-PRB62:1568} is not taken into
account in the model. It is known that the tunneling rate depends
weakly on temperature. However, the thermal escape rate increases
exponentially with temperature. For example, at 50 K, the thermal
escape rate of carriers occupying QDs with high energy levels (\textit{i.e.},
58 meV below $E_a$) reaches to $10^7$ s$^{-1}$. Therefore, the
tunneling transfer is disregarded in the model developed for
interpretation of temperature-dependence of luminescence of
LSE. Figure~\ref{Fig-C258}(b) presents the dependence of the FWHM on
temperature, from which, it is seen that the reduction of FWHM in the
luminescence spectra is mainly due to the effect of redistribution of
localized carriers. The effect of phonon scattering is to broaden the
spectra monotonously as the temperature increases, whereas carrier
thermal re-distribution results in a dependence showing a valley as
already mentioned earlier. The combination of the two effect leads to
the anomalous dependence of FWHM on temperature shown by the solid
curve in Fig.~\ref{Fig-C258}(b). In the calculations, the values of
$\Gamma_0$=5.0 meV, $\sigma_A$=10 $\mu$eV/K, $\gamma_{LO}$=18.8 meV
and $\hbar\omega_{LO}$=36 meV were
adopted. \cite{RudinS1990-PRB42:11218, KammererC2001-PRB65:33313}
Finally, Figure~\ref{Fig-C258}(c) gives the integrated intensity of
the luminescence. It can be found that when the recapture coefficient
$\gamma_c$=0.9 is taken, the calculated intensity agrees well with the
experimental data.

The results presented above is for $E_a-E_0>0$. The value of $E_a-E_0$
measures the mean thermal activation energy for the localized
carriers. For the system of InGaAs self-assembled QDs investigated in
this work, the delocalized state is seen to locate at an energy 73 meV
above the central position of the localized states. The origin of such
a delocalized state may lie on the presence of a wetting layer due to
the S-K growth mode of InGaAs on
GaAs.\cite{XuZY1996-PRB54:11528,XuZY1998-SM23:381}



For the case of $E_a-E_0<0$, calculated luminescence peak positions as
a function of temperature is given in Fig.~\ref{Fig-negat} for a few
selected values of $E_a-E_0$. The other parameters used are also
listed in the figure, which remain unchanged for the whole
calculations. It is noted that the ``S-shaped'' temperature dependence
of luminescence peaks \cite{ChoYH1998-APL73:1370} can be modeled
well. The solid squares stand for the experimental data for an
InGaN/GaN QWs.\cite{LiQ2004} In fact, the experimental data for
InGaN/GaN QWs samples reported by other groups
\cite{EliseevPG1997-APL71:569, CaoXA2003-APL82:3614} can be
quantitatively interpreted within the whole temperature range using
the current model. Further, for the partially ordered GaInP epilayer
system grown on GaAs, \cite{KondowM1989-APL54:1760} the observed
anomalous temperature dependence of the anti-Stokes photoluminescence
peak can also be reproduced with the current
model.\cite{XuSJ2004-APL84:2282}

Finally, we briefly discuss the physical meaning of $E_a$. Like the
Fermi level in the Fermi-Dirac distribution function, $E_a$ in the
distribution function derived in the work gives a ``marking'' level
below which all the localized states are occupied by carriers. Its
relative position to $E_0$ essentially determines anomalous
temperature dependence of luminescent peak position. For the cases of
InGaN/GaN QWs ($E_a<E_0$), $E_a$ may be the quasi-Fermi level of
samples,\cite{EliseevPG1997-APL71:569} which depends on concentration
of carriers optically/electrically injected and magnitude of the
built-in electric field within the samples. It is obvious that the
value of $E_a-E_0$ depends individually on sample. As shown in
Fig.~\ref{Fig-negat}, the magnitude of $E_a-E_0$ determines the
details of peak position variation with temperature. Another point we
want to mention is that if $E_a$ is far below $E_0$, \textit{i.e.},
$E_0-E_a\ge 8\sigma$, the model will be no longer valid. Under such
condition, the luminescent density, $\frac{N(E_a,T)}{\tau_r}\sim10^{-5}$
(photon number per second), could be too weak to observe when a
typical value of $\tau_r$=1 nanosecond was considered.



In conclusion, a model is developed, which quantitatively describes
the temperature-dependence of the luminescence spectra from localized
carriers. It reproduces almost all the anomalies observed for the
luminescence of LSE. It is demonstrated that the two well known
band-tail and luminescence quenching models are simply the
approximations of the current model under certain limiting
conditions.




The authors acknowledge P. Y. Yu for his critical reading of the
manuscript and valuable suggestions. We also thank H. Yang, Z. Y. Xu,
Z. L. Yuan for their helpful discussions. R. F. Duan's PL measurements
of QDs samples are acknowledged. This work was supported by the HK RGC
Grants (No. HKU 7036/03P; HKU 7049/04P), HKU Research Grants
(No. 10204008), and partially supported by HK RGC Grants (No. HKU
7118/02P).




\bibliographystyle{apsrev}%
\bibliography{references,thesis,III-Nitride,LiQ-pub,negative,local}%




\vspace{0.5em}\newpage

\noindent\textbf{Figure Captions}
\begin{description}

\item[FIG. 1.] Normalized distribution function and Gaussian-type
  density of states for localized carriers. Note that there can be
  two cases: $E_a-E_0>0$ and $E_a-E_0<0$.

\item[FIG. 2.] Calculated population densities of localized carriers
  $N(E,T)$, which resemble luminescence spectra, as a function of
  energy and temperature for the case of $E_a-E_0>0$. The curves
  are shifted along vertical direction for clarity.

\item[FIG. 3.] Calculated temperature dependence of luminescence peak
  position (a); FWHM (b); and integrated intensity (c) for the case of
  $E_a-E_0>0$\ using the parameters given for Fig.~\ref{Fig-N-ET}. The
  squares are experimental data and the solid lines are calculated
  using corresponding equations as denoted.

\item[FIG. 4.] Calculated temperature dependence of luminescence peak
  position for the cases of $E_a-E_0<0$, depicting the ``S-shape''
  dependence curve. The squares are experimental data, the solid lines
  are calculated by using Eqs.~(\ref{Energy-Varshni}) and
  (\ref{sol-x}) by setting different values of $E_a-E_0$ as denoted.

\end{description}




\newcommand{\mywidthA}{0.7\linewidth}\newcommand{\myheightA}{0.9\linewidth}

\newpage

\begin{figure}[htbp]%
  \centering\includegraphics[width=\mywidthA]{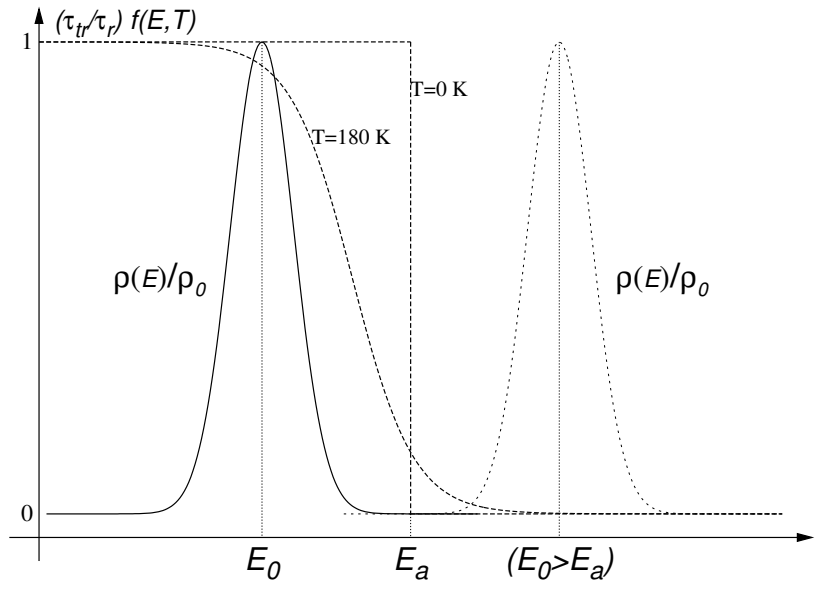}%
  \caption{of 4. Q.~Li, \textit{et al.}}%
  \label{Fig-dist-funs}%
\end{figure}%

\newpage

\begin{figure}[htbp]%
  \centering\includegraphics[height=\myheightA]{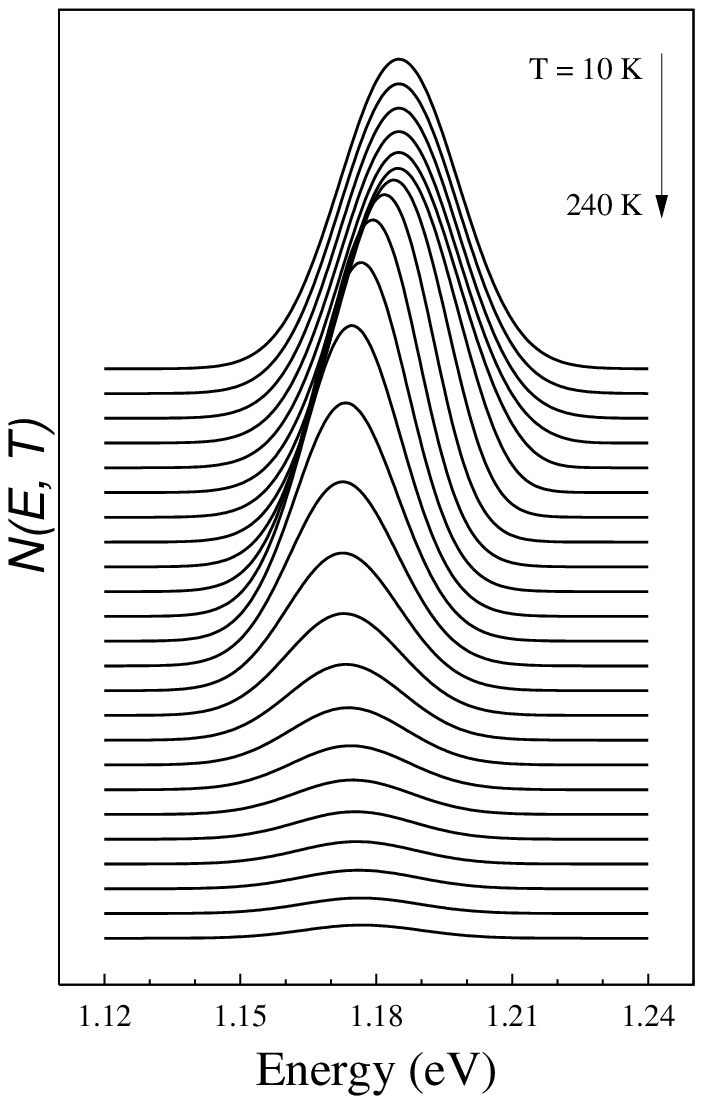}
  \caption{of 4. Q.~Li, \textit{et al.}}%
  \label{Fig-N-ET}%
\end{figure}%

\newpage

\begin{figure}[htbp]%
  \centering\includegraphics[height=\myheightA]{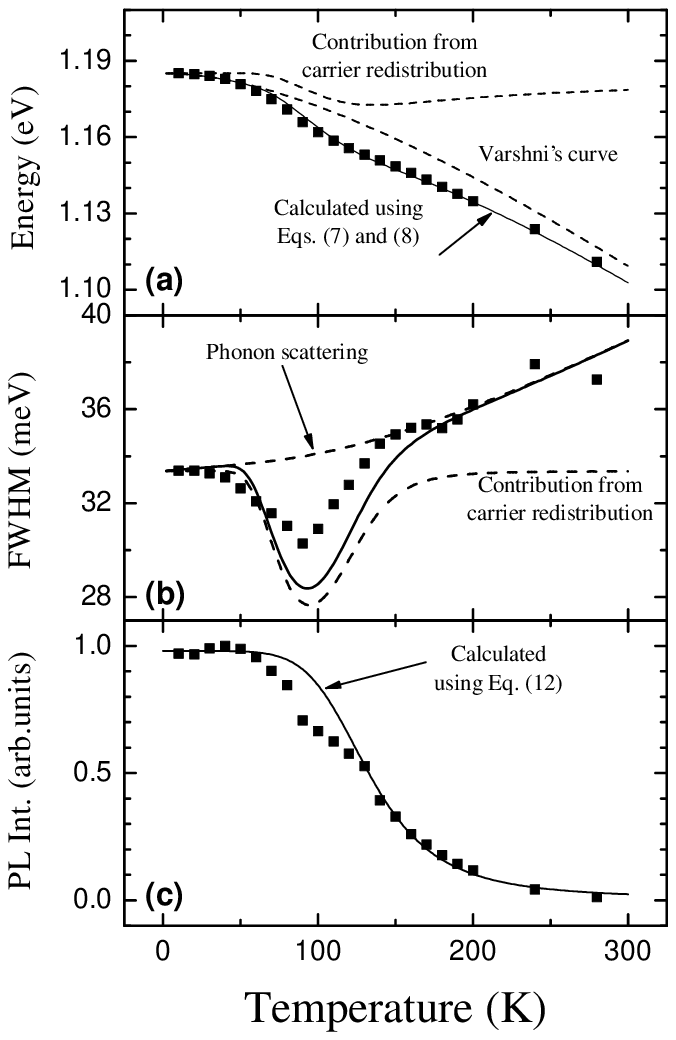}%
  \caption{of 4. Q.~Li, \textit{et al.}}%
  \label{Fig-C258}%
\end{figure}%

\newpage

\begin{figure}[htbp]%
  \centering\includegraphics[width=\mywidthA]{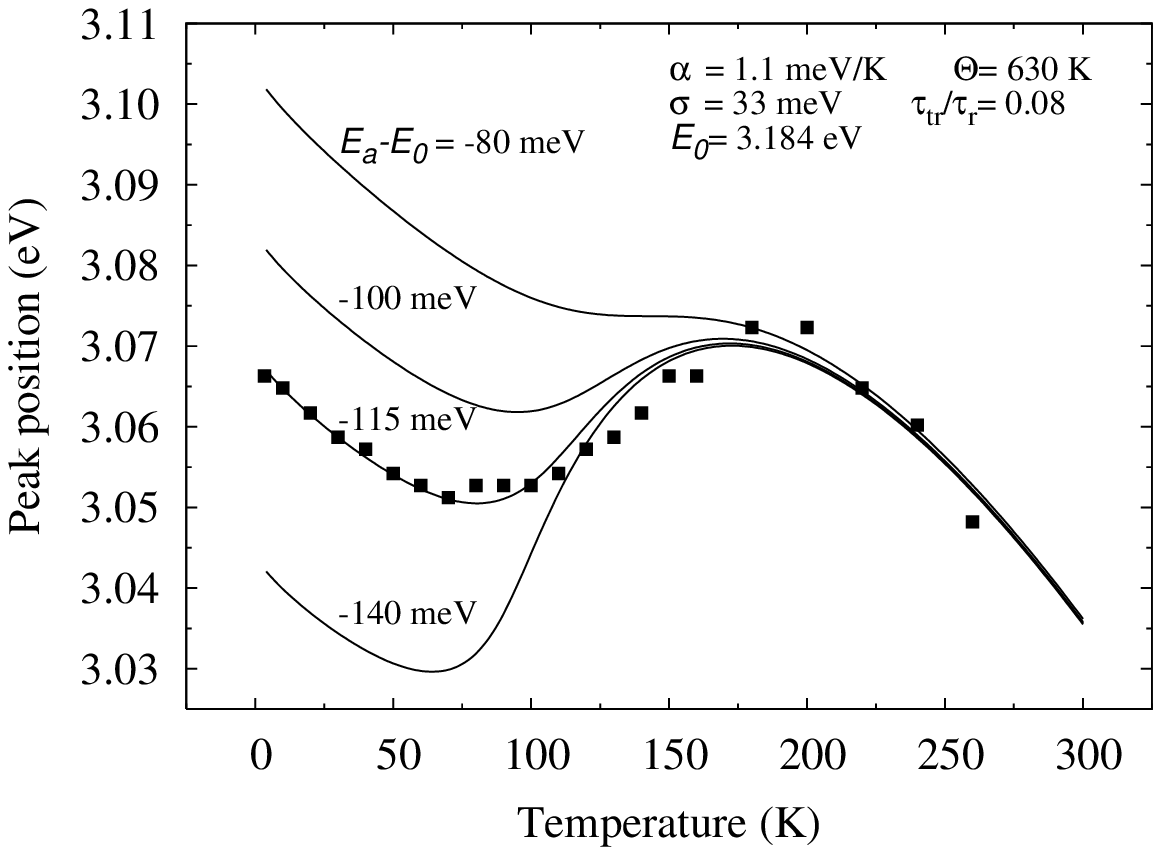}%
  \caption{of 4. Q.~Li, \textit{et al.}}%
  \label{Fig-negat}%
\end{figure}%


\end{document}